\documentclass[runningheads]{llncs}
\usepackage[T1]{fontenc}
\usepackage{graphicx}
\usepackage{hyperref}
\usepackage{hyperxmp}
\usepackage{tabularx} 
\usepackage{makecell}
\usepackage{arydshln}
\usepackage{caption}
\usepackage{subcaption}
\usepackage{amssymb}

\usepackage{etoolbox} 
\begin{document}
%
\title{State-of-the-Art HCI for Dementia Care: A Scoping Review of Recent Technological Advances}
%
%
\author{
Yong Ma\inst{1}\orcidID{0000-0002-8398-4118} \and
Yuchong Zhang\inst{2}\orcidID{0000-0003-1804-6296} \and
Oda Elise Nordberg\inst{1}\orcidID{0000-0003-3680-5524} \and
Arvid Rongve\inst{1,3}\orcidID{0000-0002-0476-4134} \and
Miroslav Bachinski\inst{1}\orcidID{0000-0002-2245-3700} \and
Morten Fjeld\inst{1,4}\orcidID{0000-0002-9562-5147}
}

\authorrunning{Y. Ma et al.}
%
\institute{University of Bergen, Bergen, Norway 
\and KTH Royal Institute of Technology, Stockholm, Sweden \and Helse Fonna, Haugesund, Noway \and Chalmers University of Technology, Gothenburg, Sweden
\\}

\maketitle              
\begin{abstract}
Dementia significantly impacts cognitive, behavioral, and functional abilities, creating challenges for both individuals and caregivers. Recent advancements in HCI have introduced innovative technological solutions to support people with dementia (PwD) and their caregivers. This scoping review systematically examines 32 recent publications from leading digital libraries, categorizing technological interventions into four key domains: Assistive and Smart Technology for Daily Life, Social Interaction and Communication, Well-being and Psychological Support, and Caregiver Support and Training. Our analysis highlights how emerging technologies are transforming dementia care. These technologies enhance quality of life by promoting independence, fostering social engagement, and providing emotional and cognitive support. However, the review also identifies critical gaps, particularly in addressing the needs of individuals with early-stage dementia and the lack of individualized support mechanisms. By emphasizing user-centered design, accessibility, and ethical considerations, this paper offers a structured roadmap for future research and practice in dementia care. It bridges the gap between technological innovation and the real-world needs of PwD and their caregivers, providing valuable insights for researchers, practitioners, and policymakers. This review not only synthesizes current advancements but also sets the stage for future HCI-driven innovations in dementia care, aiming to improve outcomes for an aging global population.
\end{abstract}

\section{Introduction}

Dementia, a progressive neurological condition characterized by cognitive decline, behavioral changes, and functional impairments, represents a significant and growing global health challenge~\cite{dening2015dementia}. As the global population ages, the prevalence of dementia continues to rise, placing increasing demands on healthcare systems and caregivers. The gradual loss of memory, cognitive abilities, and emotional and social functioning profoundly impacts the daily lives of people with dementia (PwD) and their informal caregivers, significantly diminishing their overall quality of life~\cite{potkin2002abc}. While advancing age is a well-established risk factor, it is critical to recognize that dementia is not an inevitable consequence of aging~\cite{dening2015dementia}. Instead, it is a complex and multifaceted condition that demands innovative approaches to early detection, care, and support.

Recent advancements in artificial intelligence (AI) and wearable technologies have revolutionized the early-stage diagnosis of dementia. By analyzing diverse patient data, such as brain imaging, speech patterns, facial expressions, movement, and sleep behavior, these technologies offer promising tools for identifying dementia in its initial stages~\cite{li2022applications}. However, the challenges of dementia care extend far beyond diagnosis. Providing effective care for PwD involves addressing their physical, emotional, and social needs while also supporting their caregivers, who often lack the necessary training and resources. Informal caregivers, typically family members, face significant emotional and financial burdens, underscoring the urgent need for accessible support systems and innovative solutions.

Assistive technologies, including monitoring systems, cognitive therapies, and daily living support tools, have demonstrated significant potential in enhancing dementia care~\cite{pappada2021assistive}. For instance, wearable devices equipped with advanced sensors enable continuous monitoring of daily activities, providing valuable insights into the behavior and care patterns of PwD~\cite{yang2021multimodal}. Mobile health applications extend healthcare support, improving the well-being of PwD~\cite{yousaf2019mobile}, while interactive entertainment systems foster engagement and independence, enhancing their quality of life~\cite{alm2007interactive}. However, the integration of these technologies into dementia care raises important ethical considerations, particularly regarding the stigmatization and disempowerment of PwD. Researchers in human-computer interaction (HCI) emphasize the need for tailored technologies that prioritize the voices and preferences of PwD, rather than viewing dementia solely through a medical lens as a problem to be "solved"~\cite{lazar2017critical}.

While existing review papers, such as those by Dada et al.~\cite{dada2021intelligent} and Koo et al.~\cite{koo2019examining}, have explored smart health technologies and mobile solutions for dementia care, they often focus on specific aspects of technology, such as monitoring or caregiver support. In contrast, this paper adopts a broader perspective, synthesizing advancements across multiple dimensions of dementia care technologies over the past five years (2020–2024). By categorizing these technologies into four thematic areas— Assistive and Smart Technology for Daily Life, Social Interaction and Communication, Well-being and Psychological Support, and Caregiver Support and Training - being and Psychological Support, and Caregiver Support and Training—we aim to provide a comprehensive roadmap for researchers and practitioners in the field.

Furthermore, this review emphasizes the ethical and human-centered considerations in developing technologies for PwD, aligning with the critical dementia perspective~\cite{lazar2017critical}. Unlike previous reviews that primarily focus on technological capabilities, this paper highlights the importance of tailoring solutions to the unique needs and preferences of PwD and their caregivers. By systematically mapping the existing literature from authoritative sources such as PubMed, Web of Science (WOS), Scopus, ACM, and IEEE, we offer a nuanced understanding of how technology can bridge gaps in dementia care while addressing the ethical and practical challenges associated with its implementation.

This review paper contributes to the growing body of knowledge on dementia care technologies by providing a holistic and up-to-date synthesis of advancements in the field. It not only builds on existing research but also identifies emerging trends and gaps, offering insights for future innovation. By focusing on both technological advancements and ethical considerations, this paper aims to enrich the experiences of PwD and their caregivers, ultimately improving the quality of dementia care in an aging world.
\section{Method}
Our literature review follows the PICO (Patients, Intervention, Comparison, and Outcome) framework~\cite{eriksen2018impact}, with a focus on individuals with dementia (PwD) and their caregivers. We conducted our search within the PubMed, Web of Science (WOS), Scopus, ACM, and IEEE Digital Libraries, covering research published between 2020 and 2024 to capture the most recent advancements. The Preferred Reporting Items for Systematic Reviews (PRISMA) guidelines~\cite{selccuk2019guide} informed our process, which consisted of four stages: identification, screening, eligibility, and inclusion. Figure~\ref{fig:search_flowchart} illustrates this selection process, detailing the criteria applied at each stage.
In this section, we formulate two research questions to guide our review process. These questions serve as the foundation for our systematic literature search, ensuring alignment with the key aspects of technology in dementia care. Subsequently, we outline our approach to the publication search process, illustrating how it corresponds to these research questions.

\subsection{Research Question}
As previously mentioned, dementia care aims to enhance the lives of individuals with dementia and the well-being of their caregivers. To explore how ongoing technologies are employed in the field of dementia care, we formulated the following two research questions:

\begin{figure*}[!ht]
\centering
  \includegraphics[width=.9\linewidth]{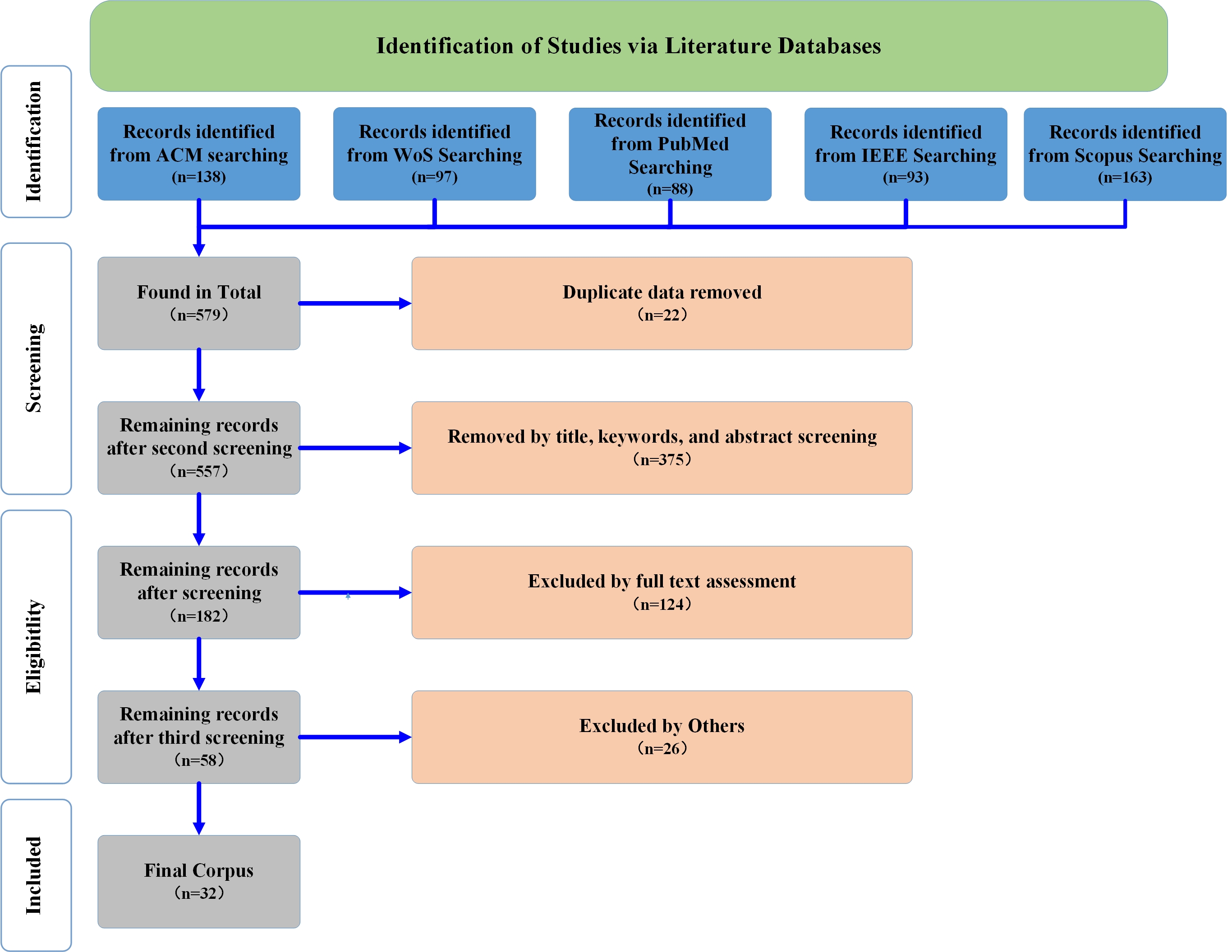}
  \caption{\scriptsize{Overview of review methodology presented on the PRISMA flow chart.}}
  \label{fig:search_flowchart}
\end{figure*}

\begin{itemize}
    \footnotesize\item \textbf{RQ1:} Which technology-based dementia care strategies can effectively improve the quality of life for people living with dementia?
    \item \textbf{RQ2:} Which technology-based stress-coping approaches can significantly alleviate the strain and distress experienced by caregivers of people living with dementia?
\end{itemize}

The integration of technology-based dementia care with traditional approaches has the potential to significantly enhance the quality of life for PwD. Emerging technologies, such as AI~\cite{kameyama2024applications}, wearable devices~\cite{lin2024implementation}, and robotics~\cite{koh2024bridging}, hold promise in revolutionizing dementia care and support. To address RQ1, we explored keywords such as "technology-based," "dementia care," "older people with dementia," and "quality of life" in our literature searches. These keywords helped us uncover a wealth of knowledge to inform our quest to optimize the well-being of individuals living with dementia.

Additionally, our exploration of current HCI research for dementia care extends to technologies that provide essential assistance to caregivers. These caregivers often bear significant emotional and financial burdens, which can lead to stress and strain. To address this, integrating technology-based solutions into their care routines can offer much-needed support. Thus, to address RQ2, we explored keywords such as "dementia caregivers" and "technology-based" to identify innovative ways of supporting and empowering these caregivers.

\subsection{Search Strategy}
\label{search_strategy}
In our literature research strategy, we employed the PICO search framework~\cite{eriksen2018impact}, which enabled us to structure our search in a comprehensive and systematic manner. The PICO model organizes keywords into four distinct categories—Person/Population, Intervention, Comparison, and Outcome—ensuring a focused and thorough exploration of the relevant literature.

\begin{itemize}
    \item[$\blacksquare$] \textbf{Person/Population (P):} This category included keywords related to the elderly population, such as "\textbf{Elderly}", "\textbf{Seniors}", "\textbf{Older Adults}", "\textbf{Older People}", "\textbf{Older Population}", "\textbf{Geriatric}", and "\textbf{Aging Population}". These terms ensured that our review captured literature specifically addressing the dementia healthcare needs of older adults, particularly those living with dementia.

    \item[$\blacksquare$] \textbf{Intervention/Indicator (I):} This category focused on keywords related to interactive technologies. We used terms such as "\textbf{Technology-based interventions}", "\textbf{Assistive technology}", "\textbf{Digital health tools}", "\textbf{Smart home technology}", "\textbf{Wearable devices}", "\textbf{Telehealth in dementia care}", "\textbf{Artificial intelligence in dementia}", "\textbf{Computer Vision}", "\textbf{Mobile apps for dementia}", "\textbf{Virtual reality dementia therapy}", "\textbf{Robotic}", "\textbf{Mobile app}", "\textbf{Virtual reality}", "\textbf{Telehealth}", "\textbf{Interactive Technology}", "\textbf{Digital Health Technologies}", and "\textbf{Robot*}". These keywords allowed us to capture the wide range of interactive applications designed to support dementia care settings.
    
    \item[$\blacksquare$] \textbf{Comparison/Control/Context (C):} To contextualize the role of interactive technologies within healthcare, we used terms such as "\textbf{dementia care}" and "\textbf{dementia healthcare}". These keywords ensured that our review remained focused on how interactive technologies impact dementia healthcare delivery, care quality, and outcomes for the elderly.

    \item[$\blacksquare$] \textbf{Outcome (O):} This category focused on keywords related to the enhancement of elderly healthcare. We used terms like "\textbf{Enhanced Healthcare}", "\textbf{Healthcare Improvement}", "\textbf{Health Enhancement}", "\textbf{Improved Health Outcomes}", "\textbf{Better Quality of Life}", "\textbf{Effective Health Monitoring}", and "\textbf{Cognitive Support}". These terms helped us identify studies that explored how interactive technologies contribute to improving health outcomes and quality of life for PwD.
\end{itemize}

By organizing our search terms into these four categories, the PICO framework allowed us to conduct a highly focused and efficient literature search. This approach ensured that our review systematically addressed the research questions by covering the relevant population (PwD), interventions (interactive technologies), comparisons (dementia care contexts), and outcomes (improved healthcare and quality of life). As a result, the PICO strategy played a pivotal role in shaping the scope and direction of our literature review, enabling us to identify and analyze the most relevant studies in the field.

\subsection{Literature Selection}

Our aim is to provide an up-to-date and comprehensive analysis of recent advancements in the field of interacitve technologies within the context of dementia care, with a particular focus on research published over the past five years (2020–2023). To achieve this objective, we employed a scoping review methodology, a robust approach that enables us to synthesize research findings, map the scope and nature of emerging research areas, and identify potential avenues for future investigation~\cite{tricco2016scoping}.

Throughout our literature collection and selection process, we adhered to the Preferred Reporting Items for Systematic Reviews (PRISMA) guidelines~\cite{selccuk2019guide} and the Meta-Analyses Extension for Scoping Reviews (PRISMA-ScR) framework~\cite{page2017evaluations}, both widely recognized methodologies for constructing comprehensive review papers. These frameworks ensured a rigorous, transparent, and systematic approach to our review process.

As illustrated in Figure~\ref{fig:search_flowchart}, our literature selection process unfolded across four iterative stages: data collection, initial filtering, research question exclusion, and full-text screening. To commence our study, we identified pertinent search terms based on our existing knowledge and executed a systematic query across five esteemed databases—PubMed, Web of Science (WoS), Scopus, ACM Digital Library, and IEEE Xplore. These databases were selected for their extensive repositories of high-quality, peer-reviewed literature, ensuring access to the most relevant and up-to-date research in the field.

\subsubsection{Identification} 
During the identification process, we utilized the aforementioned keywords to search the PubMed, WoS, Scopus, ACM, and IEEE libraries. These repositories were chosen for their comprehensive collections of literature relevant to our research objectives. Focusing on the years 2020–2024 ensured that our search yielded the most current and relevant publications. Our search retrieved 138 publications from ACM, 97 from WOS, 88 from PubMed, 93 from IEEE, and 163 from Scopus, resulting in a total of 579 relevant pieces of literature.

\subsubsection{Screening}
To ensure the relevance of the identified sources, two authors conducted a meticulous manual screening of titles, keywords, and abstracts. This process systematically reviewed and selected articles based on the following inclusion criteria:

\begingroup
\renewcommand\labelenumi{(\theenumi)}
\begin{enumerate}
    \scriptsize\item The paper was published in English, and the full text is available.
    \item The research relates to dementia care.
    \item The study participants or target users are people living with dementia, their caregivers, or both.
    \item The paper presents technologies that can support the daily lives of people with dementia.
    \item The paper addresses either RQ1 or RQ2.
\end{enumerate}
\endgroup

Prior to screening, 22 duplicate publications were removed. During screening, we expanded our criteria to include studies involving elderly individuals living with dementia, as many papers did not explicitly mention dementia patients or caregivers. Additionally, we excluded review articles and papers that only presented initial designs, technical developments, or algorithms for dementia detection. A total of 375 pieces of literature were excluded during this stage, leaving 182 records for further evaluation.

\paragraph{Eligibility}
During the eligibility stage, the same two authors assessed the full text of the remaining papers. We excluded papers that, while mentioning dementia care, primarily focused on unrelated research directions or technical developments. We also evaluated whether the technologies or designs presented could benefit PwD or their caregivers. Short papers, such as those from CHI Late-Breaking Work, were included if they addressed RQ1 or RQ2 and involved relevant participants. A total of 124 papers were excluded during this stage, resulting in 58 records.

The final corpus consisted of 32 papers, each thoroughly reviewed and aligned with our research objectives. To ensure consistency, the authors exchanged their assigned tasks and held regular discussions to resolve any uncertainties. This collaborative approach ensured a rigorous and unbiased selection process.

\subsection{Data Extraction}

The process of data extraction from our final corpus represents a critical step in addressing our research inquiries. We systematically gathered standard information, such as authors, affiliations, titles, abstracts, publication types, and publication years, from the included publications. Delving deeper into each paper allowed us to capture more nuanced details, including study objectives, motivations, technological advancements, methodologies, and environmental settings.
Our focus remained sharp on synthesizing major insights from each study, particularly identifying the technologies employed, with a keen emphasis on addressing either RQ1, RQ2, or both. Beyond the technological aspects, our analysis extended to exploring environmental settings and caregiving techniques, such as health monitoring, robotic assistance, cognitive support, wearable technology, and more.
Furthermore, the extraction process encompassed crucial details related to study design and participant information from the 32 selected papers. The culmination of these summaries and extracted critical information positions us for a comprehensive final analysis. This analysis is poised to unveil existing dementia care technologies while concurrently informing the exploration of potential technologies for future dementia care studies.
Moreover, the final summation and analysis derived from each chosen literature source hold significant promise in validating various categorizations and deepening our understanding of the primary research areas in dementia care technologies. By establishing clear categorizations and conducting a thorough analysis, we aim to create a robust framework for exploring future avenues of research in dementia care. This framework will not only highlight current advancements but also identify gaps and opportunities for innovation, ultimately contributing to the improvement of dementia care practices and outcomes.

\section{Results}
In this section, our primary focus is to present the results of our analysis, with an emphasis on recent technological advancements in the domain of dementia care. Grounded in these contemporary technologies, we have systematically categorized our findings into six distinct and coherent themes. These thematic categories are \emph{Assistive and Smart Technology for Daily Life}, \emph{Social Interaction and Communication}, \emph{Caregiver Support and Training}, \emph{Well-being and Psychological Support}. By categorizing our findings into these four research themes, we aim to offer a structured and insightful panorama of the current technological milieu within dementia care. 

\subsection{Results of Literature Search}

\begin{figure}[!ht]
    \centering
    \begin{subfigure}[b]{0.6\textwidth}
        \centering
        \includegraphics[width=\textwidth]{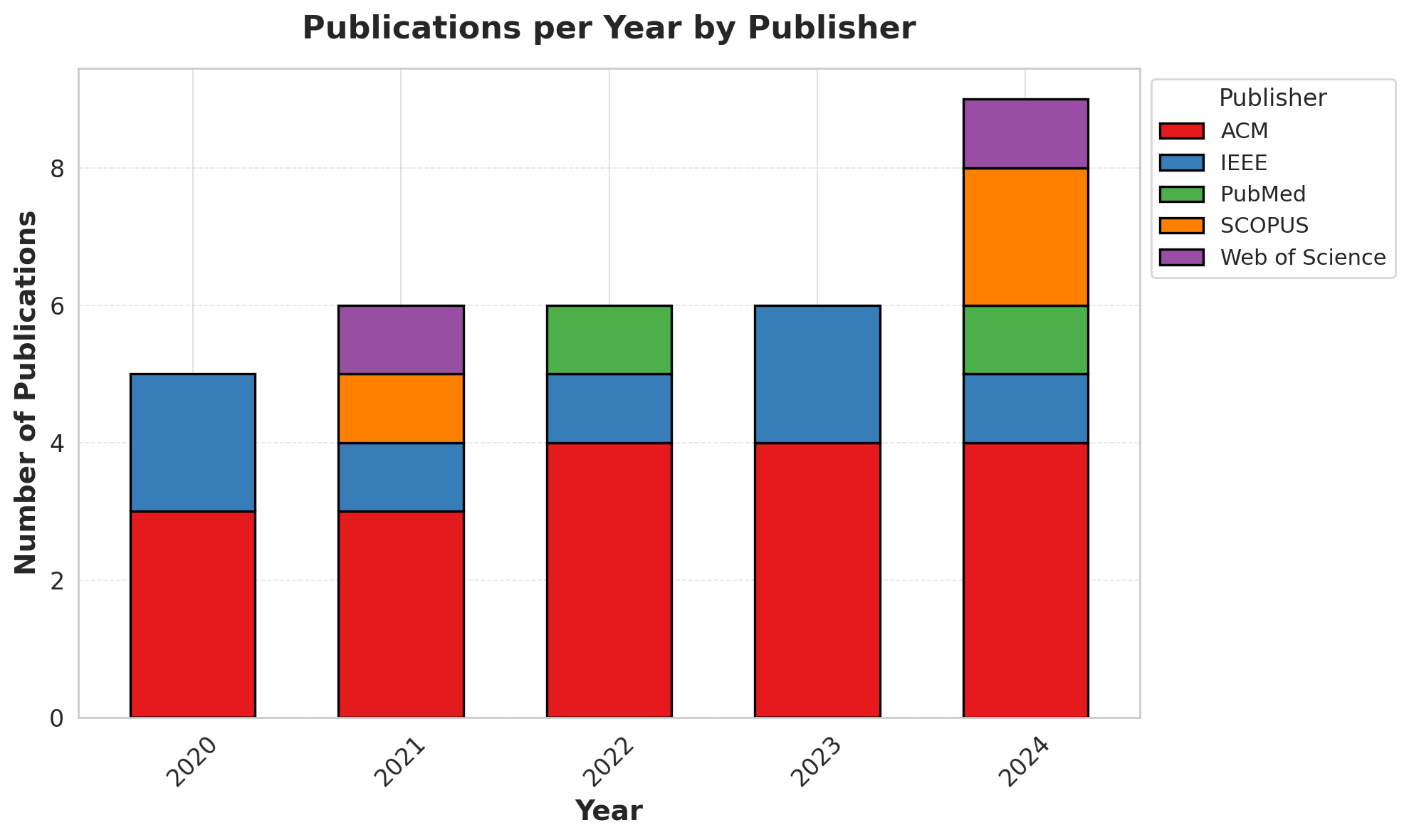}
        \caption{Number of publications by year and publisher,}
        \label{fig:year-publication}
    \end{subfigure}

    \vspace{0.5cm}
    
    \begin{subfigure}[b]{0.6\textwidth}
        \centering
        \includegraphics[width=\textwidth]{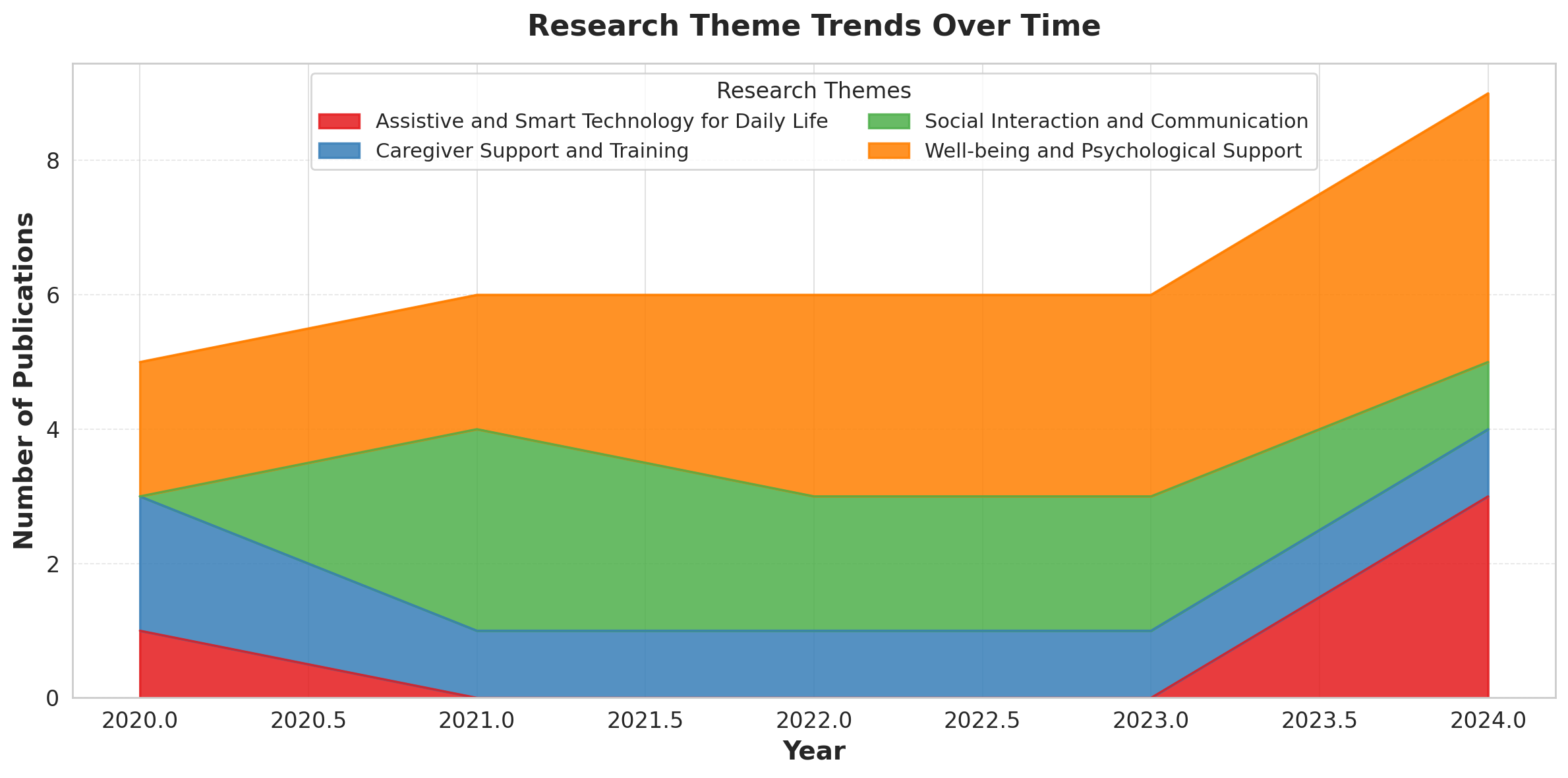}
        \caption{Trends in current technology categories in dementia care over time.}
        \label{fig:trends-themes}
    \end{subfigure}

    \vspace{0.5cm}

    \begin{subfigure}[b]{0.6\textwidth}
        \centering
        \includegraphics[width=\textwidth]{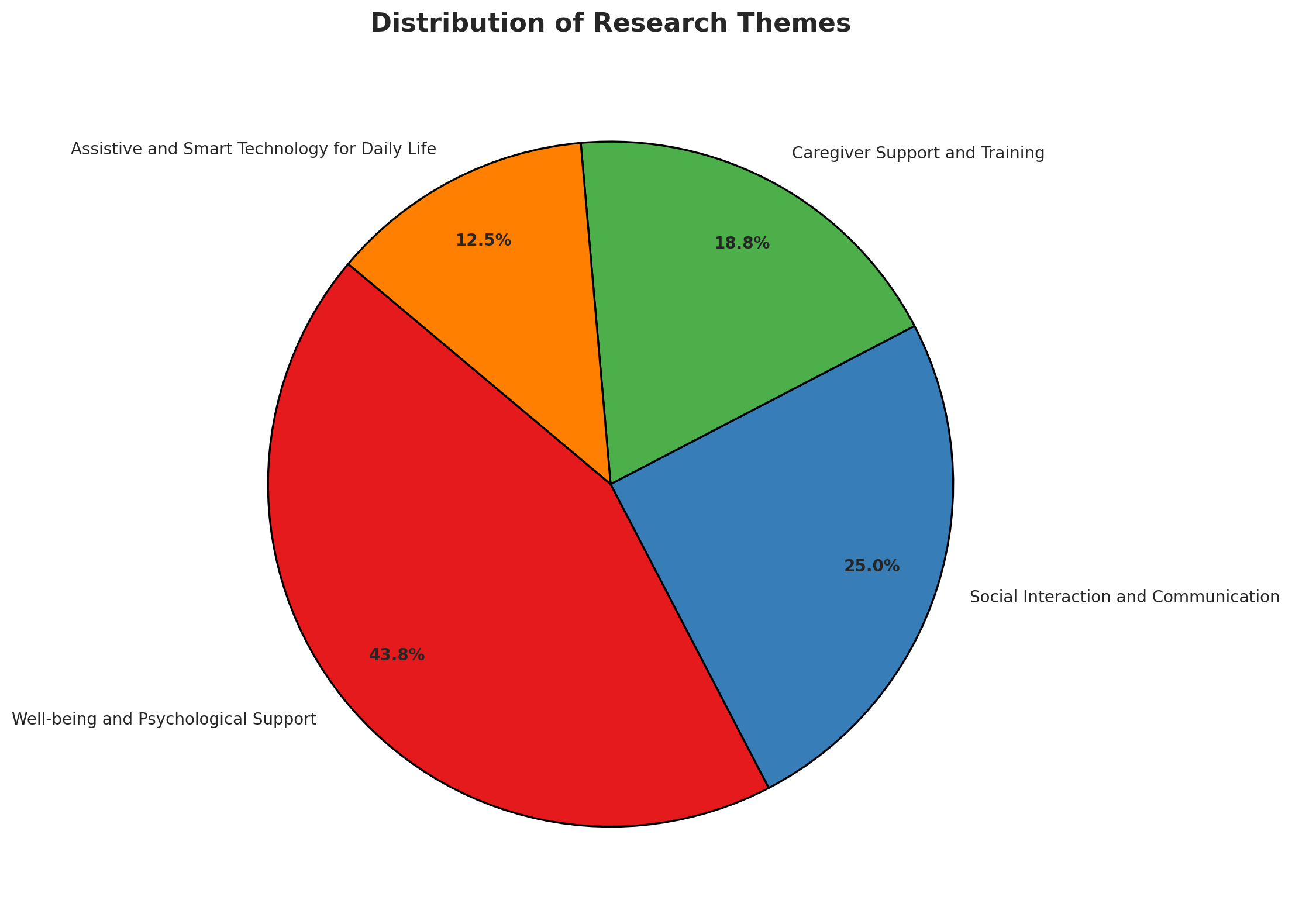}
        \caption{Distribution of current technology categories in dementia care.}
        \label{fig:topic-distribution}
    \end{subfigure}
    
    \caption{Overview of Dementia Research Publications and Themes. (a) Number of publications by year and publisher. (b) Trends in current technology categories over time. (c) Distribution of current technology categories.}
    \label{fig:dementia-research}
\end{figure}

Our research review was conducted across several comprehensive databases, including PubMed, Web of Science (WOS), Scopus, ACM, and IEEE, culminating in the selection of 32 highly relevant publications. To visually represent our findings and the evolving landscape of dementia care technology, we employed histograms, as illustrated in Figure~\ref{fig:year-publication}. This figure depicts the distribution of research publications over the years, categorized by publishers such as ACM, IEEE, PubMed, Scopus, and Web of Science. The histogram reveals a growing research interest in dementia care technologies, particularly in recent years. Notably, ACM has consistently led in the number of publications, followed closely by IEEE, highlighting the pivotal role of computing and engineering communities in advancing dementia care technologies. The inclusion of PubMed and Scopus indicates a rise in interdisciplinary collaborations, merging healthcare insights with technological innovations.

Figure~\ref{fig:trends-themes} showcases the evolution of research themes over time, focusing on key categories such as Assistive and Smart Technology for Daily Life, Caregiver Support and Training, Social Interaction and Communication, and Well-being and Psychological Support. The data reveals a steady increase in research dedicated to well-being and psychological support, underscoring the growing emphasis on emotional and cognitive interventions in dementia care. Additionally, technologies aimed at enhancing social interaction and communication have seen a progressive rise, reflecting the importance of digital and robotic solutions in fostering engagement for individuals with dementia. Fluctuations in research on caregiver support and assistive technologies suggest ongoing exploration and refinement in these areas.

Figure~\ref{fig:topic-distribution} provides a comprehensive overview of the proportion of research dedicated to various thematic areas. Well-being and psychological support dominate the landscape, emphasizing their central role in dementia care innovations. Social interaction and communication also represent a significant portion, highlighting the relevance of conversational agents, robotics, and digital engagement platforms in supporting both dementia patients and caregivers. While caregiver support and training, along with daily life assistance technologies, remain critical, they occupy a smaller share of the research, pointing to potential areas for further exploration and development. This distribution underscores the need for a holistic approach that integrates cognitive, emotional, social, and functional support in dementia care technology.

\subsection{Thematic Analysis of Dementia Care Technologies}
To systematically categorize and analyze research trends in dementia care technologies, we conducted a thematic analysis following Braun and Clarke’s~\cite{braun2006using} six-step framework. This approach enabled us to identify patterns across the selected literature and group studies based on their technological contributions. We began by thoroughly reviewing 33 selected research papers, extracting key insights regarding their objectives, methodologies, and the role of technology in dementia care. Through an iterative coding process, we identified recurring themes related to assistive technologies, social engagement, caregiver support, and well-being enhancement. Each paper was assigned multiple descriptive labels to capture its core technological focus before consolidating them into broader themes.

After the initial coding, we categorized the studies into four primary themes based on the nature of their technological intervention: (1) Assistive and Smart Technology for Daily Life, which includes solutions that aid activities of daily living (ADLs) and promote independence; (2) Social Interaction and Communication, encompassing robotic and digital tools designed to facilitate engagement and companionship for individuals with dementia; (3) Caregiver Support and Training, which includes training programs, mobile applications, and robotic assistants aimed at reducing caregiver burden and improving skill development; and (4) Well-being and Psychological Support, comprising interventions such as music therapy, reminiscence-based applications, and virtual reality experiences that focus on emotional and cognitive well-being. Each theme was refined iteratively to ensure clear distinctions while addressing overlaps in studies that spanned multiple domains.

The thematic analysis results provided valuable insights into the distribution and trends of dementia care technologies. We observed a significant focus on social interaction and well-being enhancement, reflecting the increasing importance of engagement-based interventions. The findings also highlighted a growing interest in robotic and AI-driven solutions, particularly in social and caregiver-support domains. By structuring the thematic analysis in this way, we were able to map the technological landscape, highlight key advancements, and identify research gaps in dementia care. The categorized themes serve as a foundation for discussing how technology can further evolve to meet the holistic needs of people with dementia and their caregivers.

\subsection{Review Topics of Current Technologies in Dementia Care}
Table~\ref{tab:catergories_dementia_care} provides an overview of the selected relevant papers and their corresponding categories.
Through the thematic analysis of the data material, as previously discussed, we have identified and categorized technology into four distinct areas aimed at enhancing the lives of individuals with dementia: 1) technologies for \emph{Social Interaction and Communication}, 2) technologies for \emph{Well-being and Psychological Support}, 3) technologies for \emph{Caregiver Support and Training}, 4) \emph{Assistive and Smart Technology for Daily Life}. In this section, we take a closer look at each of these categories. 

\begin{table}[htbp]
    \centering
    \scriptsize
    \caption{{The distribution of technology categorization to dementia care in the corpus.}}
    \label{tab:catergories_dementia_care}
    \begin{tabular}{ccc}
    \hline
    \makecell[c]{The Technologies-based \\ Approaches to Dementia Care} & Reference & Number \\
    \hline
    \textbf{Assistive and Smart Technology for Daily Life} & \cite{ballester2024depth,yuan2024social,mirkovic2024multimodal,carros2020exploring} & \textbf{4}\\

    \textbf{Social Interaction and Communication} & \cite{xygkou2024mindtalker,lee2023reimagining,rass2023investigating,thoolen2022livingmoments,marchetti2022pet,munoz2021evaluating,taylor2021exploring,khosla2021engagement} & \textbf{25} \\

    \textbf{Caregiver Support and Training} & \cite{wu2024effects,shen2023dementia,yuan2022robot,zubatiy2021empowering,hiramatsu2020development,lv2020teleoperation} & \textbf{6} \\
   
    \textbf{Well-being and Psychological Support} & \cite{nicol2024duet,collingham2024mariana,baumann2024mnemosyne,vidas2024family,schweiger2023robotic,houben2023switch2move,yuan2023cognitive,houben2022enriching,houben2022designing,flynn2022introducing,karaosmanoglu2021lessons,kang2021effect,houben2020role,cruz2020social} & \textbf{14} \\
     \hline
    \end{tabular}
\end{table}

\subsubsection{Technologies for Social Interaction and Communication}
Social isolation and communication difficulties are among the most significant challenges faced by individuals with dementia, leading to emotional distress, depression, and reduced cognitive function. Technologies designed for social interaction and communication aim to enhance engagement, foster meaningful relationships, and provide PwD with a sense of connectedness \cite{xygkou2024mindtalker,lee2023reimagining,rass2023investigating,thoolen2022livingmoments,marchetti2022pet,munoz2021evaluating,taylor2021exploring,khosla2021engagement}. Conversational agents, robotic companions, digital memory-sharing platforms, and cooperative gaming have emerged as effective solutions to counteract loneliness and cognitive decline. Xygkou et al. \cite{xygkou2024mindtalker} introduced AI-powered conversational agents that help PwD engage in interactive dialogues, improving their cognitive stimulation and social interaction. Similarly, Lee et al. \cite{lee2023reimagining} demonstrated how social robots could bridge intergenerational gaps by facilitating shared activities between PwD and younger family members.

In addition to robotic and AI-driven solutions, digital communication tools play an essential role in promoting social engagement and memory recall. Munoz et al. \cite{munoz2021evaluating} examined how cooperative digital games could enhance the visiting experience of PwD living in care homes, making interactions with visitors more engaging and meaningful. Similarly, memory-sharing platforms such as RelivRing \cite{van2020relivring} allow PwD to relive cherished memories through personalized audio recordings from family members, strengthening emotional bonds. Other studies highlight the importance of robotic pet therapy and virtual social platforms in alleviating social isolation, showing that digital and robotic companions can significantly improve mood, engagement, and cognitive well-being in PwD. The integration of these technologies into dementia care environments has the potential to reshape how PwD connect with their surroundings, caregivers, and loved ones, fostering greater emotional stability and quality of life.

\subsubsection{Technologies for Well-being and Psychological Support}
Ensuring the emotional well-being of individuals with dementia is just as crucial as addressing their cognitive and physical needs. Various sensory-based, music-based, and interactive cognitive therapies have been developed to enhance mood, reduce anxiety, and provide engaging experiences for PwD \cite{nicol2024duet,collingham2024mariana,baumann2024mnemosyne,vidas2024family,schweiger2023robotic,houben2023switch2move,yuan2023cognitive,houben2022enriching,houben2022designing,flynn2022introducing,karaosmanoglu2021lessons,kang2021effect,houben2020role,cruz2020social}. Nicol et al. \cite{nicol2024duet} explored music therapy interventions, demonstrating how duet-playing technology encouraged active participation and emotional expression in PwD. Similarly, Collingham et al. \cite{collingham2024mariana} developed a multi-sensory interactive platform that stimulates vision, hearing, and touch, enhancing overall emotional well-being. These approaches align with research showing that music and multi-sensory engagement can trigger positive emotions, improve social interactions, and reduce agitation in PwD.

Other studies have examined the role of virtual reality (VR), robotics, and reminiscence therapy in enhancing well-being. Baumann et al. \cite{baumann2024mnemosyne} developed VR-based reminiscence therapy, allowing PwD to relive past experiences through immersive digital environments. Schweiger et al. \cite{schweiger2023robotic} investigated robot-assisted emotion regulation, where interactive robotic pets and humanoid robots provided comfort and companionship, reducing feelings of loneliness. Additionally, sensory-based experiences, such as ambient soundscapes and AI-driven mood enhancement systems, have shown promise in stabilizing emotions and reducing stress levels. These studies highlight the need for holistic approaches in dementia care, ensuring PwD experience meaningful engagement, comfort, and joy throughout their journey.

\subsubsection{Assistive and Smart Technology for Daily Life}
Assistive and smart technologies are becoming indispensable tools in dementia care, providing individuals with dementia (PwD) greater independence in their daily activities while reducing caregiver burden. These technologies encompass wearable devices, sensor-based monitoring systems, robotic assistants, and smart home environments that help PwD navigate their daily routines safely and effectively \cite{ballester2024depth,yuan2024social,mirkovic2024multimodal,carros2020exploring}. For example, Ballester et al. \cite{ballester2024depth} introduced a depth-based interactive system designed to assist PwD with mobility challenges by detecting their presence and providing real-time guidance. Similarly, Yuan et al. \cite{yuan2024social} explored robot-assisted self-care technologies, which help PwD with dressing, grooming, and hygiene tasks, enabling them to maintain dignity and independence. The adoption of these smart technologies contributes to a safer and more structured living environment, mitigating risks such as falls, forgetting daily routines, or failing to take prescribed medications.

Beyond physical assistance, automated and AI-driven technologies are being integrated into dementia care to optimize daily life interactions. Mirkovic et al. \cite{mirkovic2024multimodal} developed a multimodal AI-driven system that adapts to the cognitive state of PwD, providing personalized reminders and interactive guidance throughout the day. Smart home environments embedded with Internet of Things (IoT) sensors are particularly useful for monitoring cognitive and physical decline over time, allowing caregivers and healthcare professionals to intervene proactively when necessary. Additionally, robotic assistants are playing an increasingly important role in promoting routine-based independence, assisting PwD in performing daily tasks while offering companionship and emotional support. The continued advancement of assistive and smart technologies holds the potential to transform home-based and institutional dementia care, enhancing quality of life while alleviating the pressures placed on caregivers and medical professionals.

\subsubsection{Technologies for Caregiver Support and Training}
Caring for individuals with dementia is a demanding task, often leading to burnout, emotional exhaustion, and high stress levels among caregivers. To address these challenges, various technology-assisted caregiver support systems have been developed, focusing on training, remote monitoring, and robotic assistance \cite{wu2024effects,shen2023dementia,yuan2022robot,zubatiy2021empowering,hiramatsu2020development,lv2020teleoperation}. Wu et al. \cite{wu2024effects} explored the effectiveness of virtual reality (VR)-based dementia training programs for professional caregivers, revealing that immersive simulations improved their understanding of dementia symptoms, patient behavior, and caregiving techniques. Similarly, Shen et al. \cite{shen2023dementia} introduced an augmented reality (AR)-based dementia education platform to provide interactive training modules, allowing caregivers to practice different caregiving scenarios in a safe and controlled virtual setting. These technologies equip caregivers with the skills and confidence needed to handle complex dementia care situations, ultimately enhancing the quality of care provided to PwD.

Beyond training, remote monitoring systems and robotic caregiver support have become essential in reducing workload and improving caregiving efficiency. Yuan et al. \cite{yuan2022robot} examined robot-assisted psychoeducation programs, demonstrating that robotic technology could provide emotional and psychological support to caregivers by engaging PwD in cognitive and physical activities. Hiramatsu et al. \cite{hiramatsu2020development} developed an AI-driven monitoring system that continuously tracks the health and safety of PwD at home, sending real-time alerts to caregivers when necessary. These advancements bridge the gap between professional and informal caregiving, offering caregivers greater flexibility, reduced stress, and improved job satisfaction. Future research should continue refining these technological caregiving solutions, ensuring they are user-friendly, scalable, and accessible to a broader caregiver population.


\section{Discussion}
Our review underscores the dynamic evolution of technological interventions in dementia care, spanning assistive and smart technologies, social interaction tools, caregiver support, and psychological well-being enhancements. The synthesis of existing literature reveals substantial progress in these areas, yet it also highlights critical gaps and opportunities for future advancements. This discussion contextualizes the findings within broader dementia care paradigms, reflecting on their implications for both individuals with dementia (PwD) and caregivers.


\subsection{Consideration of Dementia Progression}
While many technologies aim to enhance the lives of people with dementia, such as intelligent assistive technologies~\cite{dada2021intelligent}, mobile-based solutions~\cite{koo2019examining,ye2023researched}, ambient-assisted living technologies~\cite{gettel2021dementia}, and smart health technologies~\cite{guisado2019factors},etc., they often predominantly focus on individuals with later onset dementia, typically diagnosed after the age of 65. This focus leaves a notable gap in research and care for those with early onset dementia, also known as young onset dementia. Early onset dementia affects individuals who are diagnosed at a younger age, and addressing their specific needs and challenges is of paramount importance \cite{yuan2023cognitive}. 
Due to its rarity, dementia in young individuals is frequently misdiagnosed or ignored, in contrast to older adults~\cite{sullivan2022peer}. Early diagnosis is essential for obtaining appropriate support services and therapies.
Younger individuals may have different care needs compared to older adults with dementia~\cite{hancock2006needs}. They may require services tailored to their age group, such as vocational rehabilitation, educational support, and assistance with maintaining social connections. 
Furthermore, young individuals with dementia frequently want to preserve their independence and sense of self for as long as feasible~\cite{greenwood2016experiences}. Support services should focus on keeping people active and involved in things they love, as well as assisting them with difficult chores.
Additionally, unlike seniors, young people with dementia are more likely to be employed, have dependent children, and be physically active. Consequently,  the stress of dealing with dementia tends to affect younger people more than older individuals due to career and financial concerns~\cite{greenwood2016experiences}, as well as family and caregiver responsibilities~\cite{cabote2015family}. 
Moreover, younger people with dementia may experience social isolation and stigma because of their age, and support groups and social activities created for older individuals may not be appropriate for them. In terms of health needs, young-onset dementia may have distinct underlying causes and progressions than dementia in older people, therefore health management and care measures should be tailored accordingly.
However, some researchers have underscored the importance of addressing the unique requirements of individuals with early onset dementia and advocated for the implementation of dementia care technologies tailored to this demographic \cite{yuan2023cognitive}. The distinct experiences and challenges faced by individuals diagnosed at a younger age necessitate innovative approaches to technology-driven care and support.

Moreover, it is essential to recognize that dementia manifests across various stages, each of which presents different cognitive, physical, and emotional challenges. The progression of dementia can significantly influences how individuals interact with technology, and this aspect deserves further exploration. While technology has the potential to improve the quality of life for PwD at different stages of dementia, it should be designed with an understanding of the unique needs and capabilities associated with each stage.
Incorporating technology into dementia care for early onset cases and tailoring solutions to different stages of dementia will not only address existing gaps in research but also contribute to more effective and compassionate care for individuals across the entire dementia spectrum.
Future research should prioritize the development of technologies designed specifically for younger individuals with early-onset dementia, addressing challenges such as work-life balance and family responsibilities. Additionally, adaptive systems that respond to varying cognitive stages could significantly enhance the utility of technology across the entire dementia spectrum.


\subsection{Emerging Technologies and Future Trends}

Research in the field of HCI concerning technology for PwD has primarily emphasized enhancing well-being in recent years, as shown in Figure~\ref{tab:catergories_dementia_care}. This underscores the growing importance of technologies aimed at improving the quality of life for both dementia patients and their caregivers.
In addition to conventional dementia care approaches, innovative HCI technologies like virtual reality and mindfulness applications are emerging as valuable tools to enhance the well-being of individuals with dementia. These technologies offer promising avenues for improving the overall quality of life for this demographic.
The recent advancements in sensors and wearable technologies have brought about significant developments in both the \textit{Daily Life Monitoring} and \textit{Daily Life Support} categories. Smart homes and assistive devices \cite{lazarou2016novel}, among other solutions, are simplifying daily routines for dementia patients while simultaneously alleviating the burden on caregivers. These practical aids hold the potential to greatly enhance the lives of both patients and their caregivers.
Moreover, the continuous progress in large language models and natural language-based technologies suggests a potential increase in research focused on social interactions and communication. These innovations may not only bolster social robots and communication apps but also extend their application to other categories detailed in Figure~\ref{tab:catergories_dementia_care}. These technologies can offer companionship and emotional support to individuals with dementia, ultimately reducing the physical and emotional strains on caregivers. 
For instance, a personalized voice assistant tailored for dementia patients and caregivers ~\cite{li2020personalized} and social assistant robots designed for people with dementia~\cite{striegl2021designing} offer valuable assistance in managing the dementia and enhancing overall well-being.

Currently, the ongoing research in dementia care technology holds enormous promise for improving how we support individuals living with dementia. 
For instance, the innovative Rewind system leverages self-tracked location data to aid in recalling everyday memories~\cite{tan2018rewind}. While currently not implemented in dementia care, the potential benefits for future applications in this field are promising.
Additionally, NeuralGait represents another noteworthy smartphone-based system designed to passively capture gait data for assessing brain health~\cite{li2023neuralgait}. Its potential application in future dementia care holds significant promise.
Moreover, advancements in technology and social media, embedded into products we use daily such as smartphones and voice assistants, can also be applied to future dementia care~\cite{shu2021use}.
By addressing their unique needs and those of their caregivers, these technologies have the potential to revolutionize dementia care and improve the well-being of those affected by this condition. As we continue to advance in this field, it is crucial to maintain a strong focus on ethical considerations, longitudinal studies, and real-world implementation to ensure that these technologies deliver on their promise and truly enhance the lives of dementia patients and their caregivers.

\subsection{Limitation}
Despite the valuable insights gained from this systematic review, several limitations should be acknowledged. First, the scope of reviewed studies was constrained to publications from major academic databases, such as PubMed, Web of Science, Scopus, ACM, and IEEE, potentially leading to the exclusion of relevant but unpublished or non-indexed research. This may introduce a publication bias, favoring studies with positive findings over those with inconclusive or negative results~\cite{wu2024effects}. Additionally, while this review categorizes dementia-care technologies into distinct themes—such as social interaction, assistive technologies, caregiver support, and well-being enhancement—the boundaries between these categories are sometimes fluid, leading to potential overlaps in classification~\cite{shen2023dementia,baumann2024mnemosyne}.

Another limitation concerns the diversity of methodologies used across studies, making direct comparisons challenging. The heterogeneity in study designs, participant demographics, and evaluation metrics limits the generalizability of findings, particularly regarding the long-term efficacy of interventions~\cite{nicol2024duet,hiramatsu2020development}. Additionally, most studies were conducted in controlled environments, such as care facilities or hospitals, rather than real-world home settings, where external factors could influence the effectiveness of assistive technologies~\cite{lv2020teleoperation}. Future research should focus on longitudinal studies to evaluate the sustained impact of these technologies in diverse settings and across various dementia stages~\cite{thoolen2022livingmoments,khosla2021engagement}. Addressing these limitations will enhance the applicability and scalability of technology-driven dementia care solutions.

\section{Conclusion}
In this work, we have explored the landscape of dementia care technologies by analyzing literature published in PubMed, Web of Science (WOS), Scopus, ACM, and IEEE from 2020 to 2024. Our literature review approach, guided by the PICO process and PRISMA guidelines, led us to a final selection of 32 relevant papers. These papers shed light on the current state of technologies designed to enhance dementia care. Our exploration of these selected papers allowed us to categorize existing dementia care technologies into four distinct categories: \textit{Assistive and Smart Technology for Daily Life}, \textit{Social Interaction and Communication }, \textit{Caregiver Support and Training}, \textit{Well-being and Psychological Support}. Each category represents a vital aspect of dementia care, addressing the diverse needs of both patients and caregivers, while also providing answers to our aforementioned research questions.
Furthermore, we have provided a glimpse into the research agenda for each of these categories, showcasing the ongoing advancements and innovations in the field. Wearable devices, mobile applications, robotic technologies, and more have been at the forefront of research, offering promising solutions to improve the lives of those affected by dementia. Looking ahead, we have discussed potential future technologies that could revolutionize dementia care. These technologies, ranging from wearable devices to Human-Computer Interaction (HCI)-driven solutions, hold the promise of enhancing the quality of life for dementia patients and alleviating the burdens faced by caregivers.
In conclusion, our work serves as a valuable guide for future research in dementia care. By synthesizing the wealth of knowledge within this critical domain, we hope to inspire and inform researchers, caregivers, and healthcare professionals alike, ultimately advancing the care and support available to individuals living with dementia.

\newpage

%
%
%

\makeatletter
\patchcmd{\thebibliography}{\chapter*}{\section*}{}{}
\renewenvironment{thebibliography}[1] 
  {\scriptsize  
   \section*{\refname}%
   \list{\@biblabel{\arabic{enumiv}}}%
        {\settowidth\labelwidth{\@biblabel{#1}}%
         \leftmargin\labelwidth
         \advance\leftmargin\labelsep
         \usecounter{enumiv}%
         \let\p@enumiv\@empty
         \renewcommand\theenumiv{\arabic{enumiv}}}%
   \sloppy
   \clubpenalty4000
   \@clubpenalty \clubpenalty
   \widowpenalty4000%
   \sfcode`\.\@m}
  {\endlist}
\makeatother

\bibliographystyle{splncs04}
\bibliography{reference}

%




\end{document}